\documentclass[aip,jcp,reprint,floatfix]{revtex4-1}
\usepackage{epsfig}
\usepackage{graphicx}
\usepackage{dcolumn}
\usepackage{amsmath,amssymb}
\usepackage{pstricks}
\usepackage{color}
\usepackage{footnote}
\usepackage{bm}
\usepackage{bbold}
\usepackage[colorlinks=true,citecolor={blue},urlcolor={blue}, linkcolor=blue]{hyperref}
\begin{document}
\title{Calculation of hyperfine structure constants of small molecules using Z-vector method in the relativistic coupled-cluster framework}
\author{Sudip Sasmal}
\affiliation{Electronic Structure Theory Group, Physical Chemistry Division, CSIR-National Chemical Laboratory, Pune, 411\,008, India}
\author{Kaushik Talukdar}
\affiliation{Department of Chemistry, Indian Institute of Technology Bombay, Powai, Mumbai 400\,076, India}
\author{Malaya K. Nayak}
\affiliation{Theoretical Chemistry Section, Bhabha Atomic Research Centre, Trombay, Mumbai 400\,085, India}
\author{Nayana Vaval}
\affiliation{Electronic Structure Theory Group, Physical Chemistry Division, CSIR-National Chemical Laboratory, Pune, 411\,008, India}
\author{Sourav Pal}
\affiliation{Department of Chemistry, Indian Institute of Technology Bombay, Powai, Mumbai 400\,076, India}
\begin{abstract}
The Z-vector method in the relativistic coupled-cluster framework is employed to calculate the parallel and perpendicular
components of the magnetic hyperfine structure constant of a few small alkaline earth hydrides (BeH, MgH, and CaH) and fluorides (MgF and CaF).
We have compared our Z-vector results with the values calculated by the extended coupled-cluster (ECC) method reported in Phys. Rev. A {\bf91} 022512 (2015).
All these results are compared with the available experimental values. The Z-vector results are found to be in better agreement with the
experimental values than those of the ECC values.
\end{abstract}
\maketitle
\section{Introduction}
The precise calculation of the wavefunction of a many-electron system is extremely important to obtain accurate result
of different atomic and molecular properties. For the calculation of various properties like hyperfine structure (HFS) constants,
and various parity ($\mathcal{P}$) and
time reversal ($\mathcal{T}$) -violating interaction constants (like effective electric field ($E_{\mathrm{eff}}$) experienced
by the unpaired electron and scalar-pseudoscalar (S-PS) ${\mathcal{P,T}}$-odd interaction constant ($W_\mathrm{s}$)), an accurate
wavefunction is needed in the near nuclear region. The calculation of these types of ${\mathcal{P,T}}$-odd properties
is very important because they cannot be measured by any experimental technique and are very important to unravel
many mysteries of physics \cite{sakharov_1967, ginges_2004, sandars_1965, sandars_1967, labzovskii_1978, shapiro_1968, pospelov_2005}.
The accuracy of these types of properties can be accessed by comparing theoretically obtained
HFS constants with available experimental values. Light hetero-diatomic molecules are very useful for testing
the accuracy of wavefunction in the near nuclear region as the magnetic moment of both nuclei interacts with the unpaired
electron to produce HFS splitting \cite{knight_1972}.

The accurate calculation of HFS needs to include both relativistic and electron correlation effects in an intertwine manner.
The Dirac-Hartree-Fock (DHF) method is the most elegant way to include the relativistic effects in a single determinant theory.
The dynamic correlation of opposite spin electrons which is missed in DHF theory can be included by single reference coupled
cluster (SRCC) method \cite{cizek_1967, bartlett_1978}. The SRCC equations can be solved either by variational (VCC) or non-variational (NCC) way. In NCC,
the property calculation can be done either in the expectation value or in the energy derivative method. In general, these two methods
produce two different results as the NCC is non-variational and thus, does not satisfy the generalized Hellmann-Feynman (GHF)
theorem \cite{monkhorst_1977, sekino_1984}. However, the energy derivative method is superior to the expectation value method because
the first-order energy derivative is
the corresponding expectation value plus some additional terms, which makes the derivative approach closer to full configuration
interaction property value. The expectation value in NCC also produces a non-terminating series and one needs to use some truncation
scheme for practical purpose. This introduces an additional error.

On the other hand, VCC satisfies the GHF theorem and thus, both expectation value and derivative methods produce identical results.
Among many VCC \cite{szalay_1995}, extended coupled-cluster (ECC) \cite{arponen_ecc, bishop_ecc},
expectation value coupled-cluster (XCC) \cite{bartlett_xcc, pal1984xcc}, and unitary coupled-cluster
(UCC) \cite{kutzelnigg_ucc, pal1983use, pal1984ucc, tanaka_1984, hoffmann_1988, bartlett_ucc} are the
most familiar in literature. Recently, we implemented ECC in four-component relativistic framework to calculate the first-order
properties and we applied this to calculate the magnetic HFS constants of atoms and molecules \cite{sasmal_ecc}. In ECC, the amplitude equations for both excitation
and deexcitation operators are coupled and thus, we have to solve them simultaneously. In the Z-vector method (an energy derivative method
in NCC) \cite{schafer_1984, zvector_1989}, the equations for excitation operators are decoupled from the equations of deexcitation operators.
This accelerates the convergence in Z-vector method and saves enormous computational time.

In this paper, we have calculated parallel and perpendicular magnetic HFS constants of BeH, MgH, CaH, MgF, and CaF molecules using the Z-vector
coupled-cluster method and compared them with the experimental results wherever available. We also compared our results with the ECC results taken from
Ref. \cite{sasmal_ecc}.

The manuscript is organized as follows. A brief overview of the Z-vector method including concise details of
parallel and perpendicular components of the magnetic HFS constant is described in Sec. \ref{theory}. Computational details
are given in Sec. \ref{comp}. We presented our calculated results and discussed about those in Sec. \ref{res_dis} before making the
final remark in Sec. \ref{conc}.
Atomic unit is used consistently unless stated otherwise.
\section{Theory}\label{theory}
\subsection{Z-vector method}\label{zveccc}
In a single determinant theory, the DHF method yields the best description of the ground state and for
this reason, it is used as a reference state for the treatment of missing electron correlation.
The Dirac-Coulomb (DC) Hamiltonian is used in this calculation and can be expressed as
\begin{eqnarray}
{H_{DC}} &=&\sum_{i} \Big [-c (\vec {\alpha}\cdot \vec {\nabla})_i + (\beta -{\mathbb{1}_4}) c^{2} + V^{nuc}(r_i)+ \nonumber\\
       && \sum_{j>i} \frac{1}{r_{ij}} {\mathbb{1}_4}\Big],
\end{eqnarray}
where, {\bf$\alpha$} and $\beta$ are the usual Dirac matrices, $c$ is the speed of light,
${\mathbb{1}_4}$ is the 4$\times$4 identity matrix. $V^{nuc}(r_i)$ is the nuclear potential term
and we have used Gaussian charge distribution for this purpose.
The DHF method transforms the complicated many-electron problem into many one-electron problems
and in this process misses the correlation between opposite spin electrons. In this work, the missing
dynamic correlation is incorporated by using the SRCC method. The SRCC wavefunction is given by
\begin{eqnarray}
|\Psi_{cc}\rangle=e^{T}|\Phi_0\rangle ,
\end{eqnarray}
where $\Phi_0$ is the DHF wavefunction and $T$ is the coupled-cluster excitation operator which is given
by
\begin{eqnarray}
 T=T_1+T_2+\dots +T_N=\sum_n^N T_n ,
\end{eqnarray}
with
\begin{eqnarray}
 T_m= \frac{1}{(m!)^2} \sum_{ij\dots ab \dots} t_{ij \dots}^{ab \dots}{a_a^{\dagger}a_b^{\dagger} \dots a_j a_i} ,
\end{eqnarray}
where $i, j$ are the hole and $a, b$ are the particle indices. $t_{ij..}^{ab..}$ are the cluster amplitudes corresponding 
to the cluster operator $T_m$. The equations for n-body cluster amplitudes are given as
\begin{eqnarray}
 \langle \Phi_{i..}^{a..} | (H_Ne^T)_c | \Phi_0 \rangle = 0 ,
\label{cc_amp_gen}
\end{eqnarray}
where H$_N$ is the normal ordered DC Hamiltonian and subscript $c$ means only the connected terms exist in the
contraction between H$_N$ and T.
In the coupled-cluster single and double (CCSD) model, $T=T_1+T_2$. The explicit equations for T$_1$ and T$_2$ are
given as
\begin{eqnarray}
 \langle \Phi_{i}^{a} | (H_Ne^T)_c | \Phi_0 \rangle = 0 , \,\,
  \langle \Phi_{ij}^{ab} | (H_Ne^T)_c | \Phi_0 \rangle = 0 .
 \label{cc_amplitudes}
\end{eqnarray}
Once the T$_1$ and T$_2$ amplitudes are solved, the correlation energy can be obtained as
\begin{eqnarray}
 E^{corr} = \langle \Phi_0 | (H_Ne^T)_c | \Phi_0 \rangle .
\label{cc_energy}
\end{eqnarray}

The SRCC solved in this way is non-variational in nature and thus, the SRCC energy is not optimized with respect to
the molecular orbital coefficients (C$_M$) and the determinantal coefficients (C$_D$) in the expansion of the correlated
many-electron wavefunction for a fixed nuclear geometry \cite{monkhorst_1977}. Therefore, for the energy derivative calculation,
one needs to incorporate the derivative of energy with respect to C$_M$ and C$_D$ in addition to the derivative of
these two coefficients with respect to the external field parameter. However, the derivative terms related to C$_D$ and
C$_M$ can be integrated by solving linear equations for each coefficient where the solutions are in general, perturbation dependent.
This means, for each type of properties, one needs to solve a different set of linear equations.
However, in Z-vector method, this can be avoided and the derivative terms related to C$_D$ can be incorporated by the introduction of
a perturbation independent operator, $\Lambda$ \cite{zvector_1989}. The antisymmetrized form of the $\Lambda$ operator is given as
\begin{eqnarray}
 \Lambda=\Lambda_1+\Lambda_2+...+\Lambda_N=\sum_n^N \Lambda_n ,
\end{eqnarray}
where
\begin{eqnarray}
 \Lambda_m= \frac{1}{(m!)^2} \sum_{ij..ab..} \lambda_{ab..}^{ij..}{a_i^{\dagger}a_j^{\dagger} .. ..a_b a_a} ,
\end{eqnarray}
where $i, j$ are the hole and $a, b$ are the particle indices. $\lambda_{ab..}^{ij..}$ are the cluster amplitudes corresponding 
to the cluster operator $\Lambda_m$. From the form of the $\Lambda$ operator, it is clear that it is a de-excitation operator.
The $\Lambda$ amplitude equation is given by
\begin{eqnarray}
  \langle \Phi_0 | [\Lambda (H_Ne^T)_c]_c | \Phi_{i..}^{a..} \rangle + \langle \Phi_0 | (H_Ne^T)_c | \Phi_{i..}^{a..} \rangle \nonumber\\
 +  \langle \Phi_0 | (H_Ne^T)_c | \Phi_{int} \rangle \langle \Phi_{int} | \Lambda | \Phi_{i..}^{a..} \rangle  = 0 ,
 \end{eqnarray}
where $\Phi_{int}$ is the determinant corresponding to the intermediate excitation between $\Phi_0$
and $\Phi_{i..}^{a..}$.
In CCSD model, $\Lambda=\Lambda_1+\Lambda_2$ and the explicit equations for the amplitudes of $\Lambda_1$
and $\Lambda_2$ operators are given by
\begin{eqnarray}
\langle \Phi_0 |[\Lambda (H_Ne^T)_c]_c | \Phi_{i}^{a} \rangle + \langle \Phi_0 | (H_Ne^T)_c | \Phi_{i}^{a} \rangle = 0,
\end{eqnarray}
\begin{eqnarray}
\langle \Phi_0 |[\Lambda (H_Ne^T)_c]_c | \Phi_{ij}^{ab} \rangle + \langle \Phi_0 | (H_Ne^T)_c | \Phi_{ij}^{ab} \rangle \nonumber \\
 + \langle \Phi_0 | (H_Ne^T)_c | \Phi_{i}^{a} \rangle \langle \Phi_{i}^{a} | \Lambda | \Phi_{ij}^{ab} \rangle = 0.
\label{lambda_2}
\end{eqnarray}
Finally, the energy derivative can be given as
\begin{eqnarray}
 \Delta E' = \langle \Phi_0 | (O_Ne^T)_c | \Phi_0 \rangle + \langle \Phi_0 | [\Lambda (O_Ne^T)_c]_c | \Phi_0 \rangle ,
\end{eqnarray}
where, $O_N$ is the derivative of the normal ordered perturbed Hamiltonian with respect to the external field of perturbation.
\subsection{Magnetic hyperfine structure constants}\label{prop}
The interaction of the electromagnetic field generated by electrons with the magnetic dipole moment of the nucleus
is responsible for the magnetic HFS \cite{lindgren_book}. The magnetic vector potential ($\vec{A}$) due to the magnetic moment ($\vec{\mu_k}$)
of a nucleus is given as
\begin{eqnarray}
 \vec{A}=\frac{\vec{\mu}_k \times \vec{r}}{r^3}.
\end{eqnarray}
The perturbed HFS Hamiltonian
of an atom due to $\vec{A}$ in the Dirac theory is given by
\begin{eqnarray}
H_{hyp}= \sum_i^n \vec{\alpha}_i \cdot \vec{A_i},
\end{eqnarray}
where $n$ is the total no of electrons and $\alpha_i$ denotes the Dirac $\alpha$
matrices for the i$^{th}$ electron.
For a diatomic molecule, the z (along the molecular axis) and x or y (i.e., perpendicular to molecular axis)
projection of the expectation value of the corresponding perturbed HFS Hamiltonian gives the parallel ($A_{\|}$)
and perpendicular ($A_{\perp}$) magnetic HFS constant, respectively. The expression for $A_{\|}$ and $A_{\perp}$
can be written as
\begin{eqnarray}
A_{\|(\perp)}= \frac{\vec{\mu_k}}{I\Omega} \cdot \langle \Psi_{\Omega} | \sum_i^n
\left( \frac{\vec{\alpha}_i \times \vec{r}_i}{r_i^3} \right)_{z(x/y)} | \Psi_{\Omega(-\Omega)}  \rangle,
\label{hfs_mol}
\end{eqnarray}
where $I$ is the nuclear spin quantum number and $\Omega$ represents the z component
(along molecular axis) of the total angular momentum of the diatomic molecule.

\section{Computational details}\label{comp}
We have used the locally modified version of DIRAC10 \cite{dirac10} program package to solve the DHF Hamiltonian and to construct
the one-electron and two-electron integrals and the HFS integrals. Gaussian charge distribution is
considered as nuclear model where the nuclear parameters \cite{visscher_1997} are taken as default values of DIRAC10 program package.
The basis functions are represented in scalar basis and restricted kinetic balance (RKB) \cite{dyall_2007} condition is applied
to generate the small component basis functions from the large component basis functions. The unphysical solutions are
removed by diagonalizing the free particle Hamiltonian. This generates equal numbers of positronic and
electronic orbital. In our molecular calculations, we have used aug-cc-pCV5Z \cite{ccpcvxz_h_b-ne} for H atom and
aug-cc-pCVQZ \cite{ccpcvxz_h_b-ne} for F, Be, and Mg atoms and dyall.cv4z \cite{dyall_s} basis for Ca atom.
All the generated orbitals are taken for the correlation calculation of BeH.
The cutoff used for the correlation calculations of MgH, MgF, CaH, and CaF molecules are 10 a.u., 10 a.u., 15 a.u., and 15 a.u., respectively.
The bond-length used for the BeH, MgH, CaH, MgF, and CaF molecules are 1.343 \AA, 1.7297 \AA, 2.003 \AA, 1.750 \AA,
and 1.976 \AA, respectively \cite{bond_expt}.
\section{Results and discussion}\label{res_dis}
\begin{table*}[ht]
\caption{Parallel ($A_{\|}$) and perpendicular ($A_{\perp}$) magnetic hyperfine structure constant of molecules in MHz }
\begin{ruledtabular}
\newcommand{\mc}[3]{\multicolumn{#1}{#2}{#3}}
\begin{center}
\begin{tabular}{lrrrrrrrrrrr}
 &  & \mc{5}{c}{A$_{\|}$} & \mc{5}{c}{A$_{\perp}$}\\
\cline{3-7} \cline{8-12}\\
 & & & & & \mc{2}{c}{$\delta \%$} & & & & \mc{2}{c}{$\delta \%$}\\
 \cline{6-7} \cline{11-12}\\
Molecule & Atom & ECC    & Z-vector & Expt. \cite{quiney_2002} & ECC & Z-vector & ECC    & Z-vector & Expt. \cite{quiney_2002}  & ECC & Z-vector \\
\hline
BeH & $^{1}$H   &  204.1 & 199.5  & 201(1) \cite{knight_1972} & 1.5 &   0.8    &  185.6 & 181.0  & 190.8(3) \cite{knight_1972} & 2.8 & 5.4 \\
    & $^{9}$Be  & -200.6 & -201.3 & -208(1) \cite{knight_1972}  & 3.7 &   3.3    & -186.0 & -186.8 & -194.8(3) \cite{knight_1972} & 4.7 & 4.3 \\
MgH & $^{1}$H   &    ×   & 276.2  & 298(1) \cite{knight_1971_H} &  ×  &   7.9    &    ×   & 271.1  & 264(1) \cite{knight_1971_H}   &  ×  & 2.6 \\
    & $^{25}$Mg &    ×   & -188.1 & -226(3) \cite{knight_1971_H} &  ×  &   20.1   &    ×   & -172.0 & -218(1) \cite{knight_1971_H}  &  ×  & 26.7 \\
CaH & $^{1}$H   &  146.4 & 143.7  & 138(1) \cite{knight_1971_H} & 5.7 &   4.0    &  141.9 & 139.0  & 134(1) \cite{knight_1971_H}   & 5.6 & 3.6 \\
    & $^{43}$Ca & -321.6 & -331.2 &    ×    &  ×  &    ×     & -295.7 & -305.2 &     ×     &  ×  &  ×  \\
MgF & $^{19}$F  &  320.9 & 322.6  & 331(3) \cite{knight_1971_F} & 3.1 &   2.6    &  153.3 & 148.7  & 143(3) \cite{knight_1971_F}   & 6.7 & 3.8 \\
    & $^{25}$Mg & -282.6 & -281.8 &    ×    &  ×  &    ×     & -270.4 & -269.6 &     ×     &  ×  &  ×  \\
CaF & $^{19}$F  &    ×   & 131.7  & 149(3) \cite{knight_1971_F} &  ×  &   13.1   &    ×   & 83.8   & 106(3) \cite{knight_1971_F}   &  ×  &  26.5 \\
    & $^{43}$Ca &    ×   & -426.4 &    ×    &  ×  &    ×     &    ×   & -408.9 &     ×     &  ×  &  ×  \\
\end{tabular}
\end{center}
\end{ruledtabular}
\label{hfs_table}
\end{table*}
\begin{figure}[ht]
\centering
\begin{center}
\includegraphics[height=5cm, width=8cm]{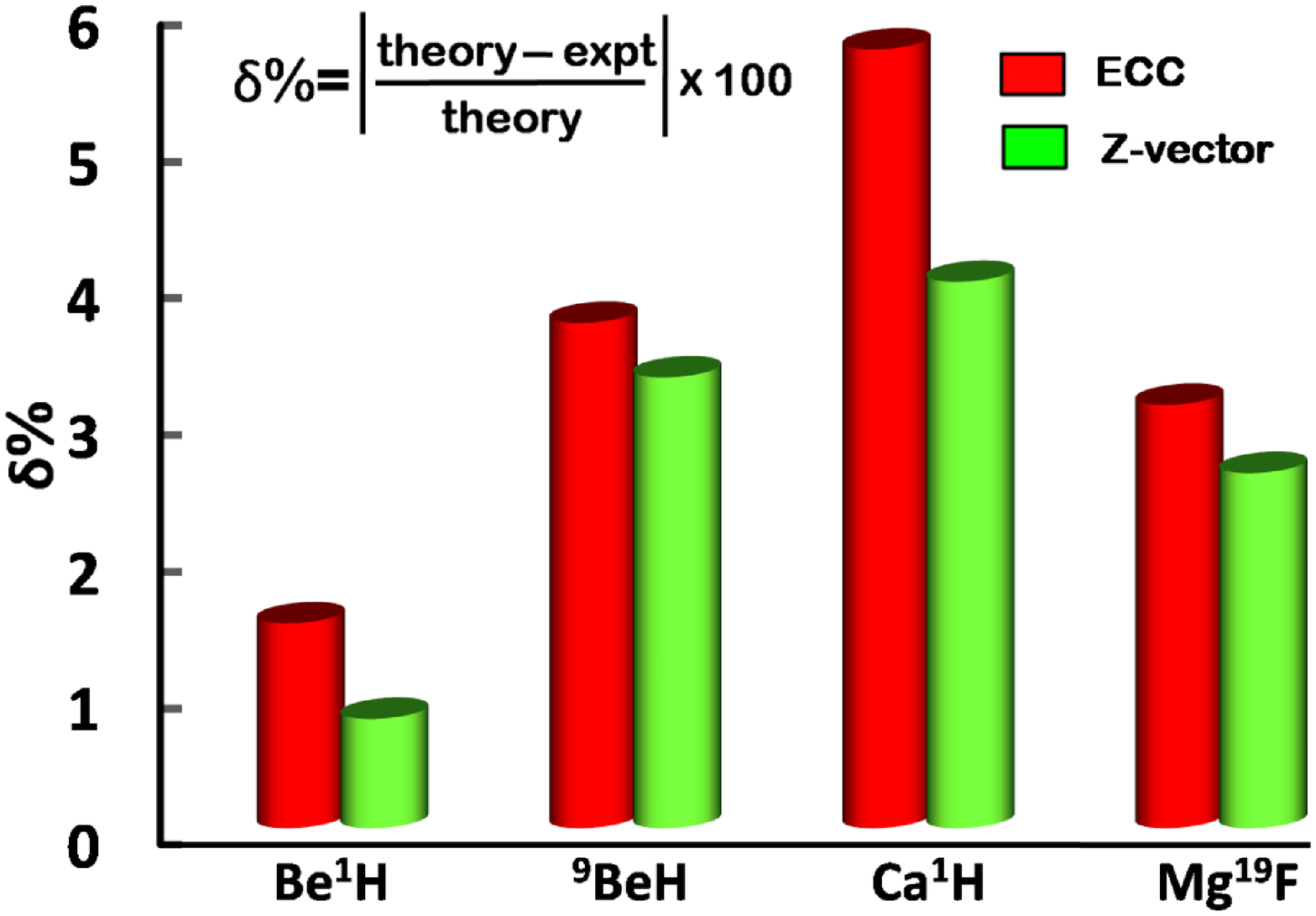}
\caption {Comparison of relative deviations of A$_{\|}$ values between ECC and Z-vector method}
\label{delta_para}
\end{center}  
\end{figure}
\begin{figure}[ht]
\centering
\begin{center}
\includegraphics[height=5cm, width=8cm]{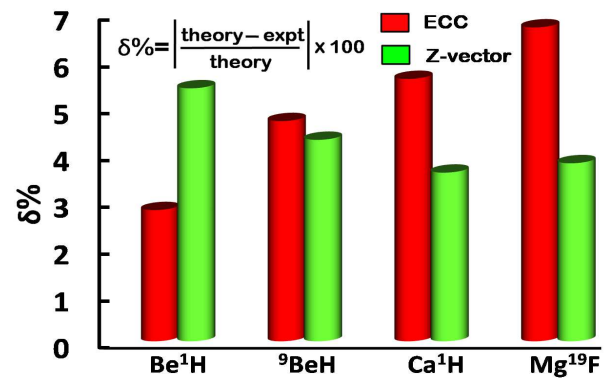}
\caption {Comparison of relative deviations of A$_{\perp}$ values between ECC and Z-vector method}
\label{delta_perp}
\end{center}  
\end{figure}
The aim of our present study is to show that the Z-vector method can produce a very good quality wavefunction
in the near nuclear region of small molecules and
for this purpose we have calculated the HFS constants of some small molecules.
In Table \ref{hfs_table}, we present the A$_{\|}$ and A$_{\perp}$ HFS constants of BeH, MgH, CaH, MgF, and CaF
using the Z-vector method in the relativistic coupled-cluster framework.
The HFS constants of BeH, MgF and CaH using the Z-vector method are also compared with the values obtained in the ECC method
reported in Ref. \cite{sasmal_ecc}.
For this reason, we have taken the identical basis and cutoff for BeH, MgF and CaH as reported in Ref. \cite{sasmal_ecc}.
These values are compared with the available experimental values \cite{quiney_2002, knight_1972, knight_1971_H, knight_1971_F}
and the deviations are presented as $\delta$\%.
From the Table \ref{hfs_table}, it is clear that our Z-vector results are in close agreement with the experimental values. The highest and lowest
deviations of Z-vector values of A$_{\|}$ occur for $^{25}$Mg of MgH and $^{1}$H of BeH, respectively, where the
deviations are 37.9 MHz and 1.5 MHz, respectively. For the Z-vector results of A$_{\perp}$, the highest and lowest
deviations appear for $^{25}$Mg of MgH and $^{1}$H of MgH, respectively, where the deviations are
46 MHz and 7 MHz, respectively.
From the above observation, we can comment that the Z-vector method can produce very good quality wavefunction in the nuclear
region of small molecules.
In Figs. \ref{delta_para} and \ref{delta_perp}, we have compared the relative deviations of A$_{\|}$ and A$_{\perp}$
values, respectively, between ECC and Z-vector method. From these figures, it is clear that the Z-vector method
always produces better results than the ECC method. The only exception is for A$_{\perp}$ value of $^{1}$H in BeH
where the difference between ECC and Z-vector values is only 4.6 MHz.
Although ECC functional generates a terminating series CCSD, the natural truncation produces computationally very costly terms.
To avoid this, the ECC functional in Ref. \cite{sasmal_ecc} is truncated in such a way that either the left vector or the
right vector is linear in a double linked form. This introduces some additional errors, which may be the reason for the poor performance
of the ECC method compared to the Z-vector method.
\section{Conclusion}\label{conc}
In conclusion, we have applied the Z-vector method in the relativistic coupled-cluster framework
to calculate the parallel and perpendicular components of the magnetic HFS constant of small diatomic
molecules. Our results show that the Z-vector can produce an accurate wavefunction in the nuclear
region of small molecules.
In Ref. \cite{sasmal_srf}, we have shown that the Z-vector method can produce
very good wavefunction in the nuclear region of moderately heavy molecules like SrF. Similarly,
Z-vector method can also produce an accurate wavefunction in the nuclear region of heavy diatomic
molecules \cite{sasmal_pbf, sasmal_hgh, sasmal_raf}. So, altogether, we can say that the Z-vector is a very reliable method
to produce high quality wavefunction in the core region.
In this work, we have also shown that the Z-vector method can produce more accurate results for the magnetic HFS constant
than the ECC method and thus, is more reliable for the calculation of various ${\mathcal{P,T}}$-odd interaction constants.
\section*{Acknowledgement}
Authors acknowledge a grant from CSIR 12th Five Year Plan project on Multi-Scale Simulations of Material (MSM)
and the resources of the Center of Excellence in Scientific Computing at CSIR-NCL. S.S and K.T. acknowledge the CSIR
for their fellowship.
S.P. acknowledges funding from J. C. Bose Fellowship grant of Department of Science and Technology (India).

\end{document}